\documentclass[12pt]{article}
\usepackage{amssymb}
\usepackage{amsmath}
\usepackage{hyperref}
\usepackage{graphicx}
\usepackage{placeins}
\usepackage{bm}
\usepackage{latexsym}
\begin{document}
\title{\bf The phase transition of Rastall AdS  black hole with cloud of strings and quintessence}
\author{Mehdi Sadeghi\thanks{Corresponding author: Email: mehdi.sadeghi@abru.ac.ir  }  \hspace{2mm} and
	Faramarz Rahmani\thanks{Email: faramarz.rahmani@abru.ac.ir }\hspace{2mm}\\
		{}\\
		{\small {\em  Department of Physics, School of Sciences,}}\\
		{\small {\em  Ayatollah Boroujerdi University, Boroujerd, Iran}}
}
\date{\today}
\maketitle

\abstract{In this paper, we introduce the black hole solution in Rastall theory of gravity in the presence of quintessence and the cloud of strings. Our investigations show that this model meets only second-order phase transition in four dimensions. While both the first and second order phase transitions are seen in five dimensions. Therefore, according to the AdS/CFT duality, the confinement-deconfinement phase transition only occurs in five dimensions for this model.}



\section{Introduction} \label{intro}

\indent Black holes behave like thermodynamic systems \cite{Kubiznak:2014zwa}. This motivates us to study the thermodynamic behavior of the black holes. The Hawking-Page phase transition which is a first-order phase transition, describes the transition between phases which are static spherically symmetric vacuum solutions of the Einstein equations in AdS spacetime. According to the gauge/gravity  duality \cite{Witten:1998qj}-\cite{Aharony}, this must corresponds to a phase transition in gauge theory. Witten shows that this correspond to a confinement-deconfinement phase transition in the gauge theory side \cite{Witten:1998zw}. The thermodynamics of black hole is a combination of general relativity and quantum field theory which helps us to formulate quantum gravity. Bardeen, Carter and Hawking were the ones who introduced the four laws of black hole thermodynamics \cite{Bardeen:1973gs}.
In this regard, the mass of a black hole and surface gravity on event horizon are interpreted as the enthalpy and the temperature of space-time respectively. This rather novel idea originates from a consideration of the Smarr relation \cite{Caldarelli:1999xj}-\cite{Smarr:1972kt}. The derivation of black hole entropy can be found in Ref \cite{Ndongmo:2021how}.\par 
Covariant derivative of the energy-momentum tensor is not zero in Rastall gravity and it is proportional to the derivative of the Ricci scalar\cite{Rastall1972}-\cite{Al-Rawaf:1996}. This may be happen in a strong gravitational field. Observational data confirm the Rastall theory of gravity. As an example, the evolution of small dark matter fluctuations is the same as that in the $\Lambda$CDM model\cite{Ndongmo:2021how}.\\
The existence of dark energy has been confirmed by Cosmic Microwave Background (CMBR) radiation \cite{WMAP:2003elm}-\cite{Planck:2013pxb} and Baryon Acoustic Oscillations (BAO) \cite{SDSS:2005xqv}. It makes up $70\%$ of the Universe.\par
 Quintessence and cloud of strings in Einstein gravity are discussed in \cite{Yin:2021akt} and we want to investigate this model for Rastall gravity\cite{Rastall:1972swe}. Rastall gravity was introduced as a modification of Einstein gravity to explain the dark matter and dark energy problem.\par 
In fact, black holes solutions surrounded by exotic matter fields have been studied to investigate the astrophysical solutions surrounded by a dark energy field in accordance with the cosmological observations about the acceleration of the universe\cite{Kiselev:2002dx}.
Rastall gravity is important from several aspects. For example, it provides an explanation for inflation and acceleration problem of the universe \cite{Zou:2020jeu}. It has been shown that the cosmological constant arises from the consistency of the non-vacuum field equations of the Rastall gravity for
a spherical symmetric spacetime\cite{Heydarzade:2016zof}. One interesting result of this model has been introduced in Ref \cite{MoraisGraca:2017hrf} which states that:\\ \textit{All electrovacuum solutions of general relativity are also solutions of Rastall gravity, but as long as (non-vanishing trace) matter is introduced in the theory, the spacetime becomes dependent of the Rastall parameter}.\\
Thus, it seems natural to examine these models in Rastall theory as well. 
In this model the dark energy component is described by both a quintessence field and a negative cosmological constant. The quintessence field is considered for the present stage, accelerating expansion, of the universe. On the other hand, the inclusion of a negative cosmological constant warrants
that the present stage of accelerating expansion will be, eventually, followed by a period
of collapse into a final cosmological singularity(AdS universe).\cite{Sadeghi:2020lfe}.\par 

In string theory, the fundamental objects are one-dimensional strings and this can be extended to a cloud of strings. Letelier\cite{Letelier:1979ej} was the one who introduced the black hole solution with the cloud of strings first time. The thermodynamics of black hole solutions in the presence of cloud of strings has been studied in Refs \cite{Ranjbari:2019ktp}-\cite{Rodrigues:2022zph}.
The accelerated expansion of the universe has been confirmed by observations \cite{WMAP:2003elm}-\cite{Planck:2013pxb}. "How the universe expanded" is a main question for cosmologists. Quintessence  model was introduced to explain this expansion by a scalar field \cite{Zlatev:1998tr}.\\
The black hole solution of Einstein’s equations coupled to the scalar fields or cosmic
strings with a global $SO(3)$ symmetry, has already been investigated in \cite{Guendelman:1991qb}. Einstein-Gauss-Bonnet AdS black hole in the presence of cloud of strings has been introduced in \cite{Herscovich:2010vr}. An investigation of the AdS black brane solution in 4-dimensional (4D) Einstein-Gauss-Bonnet-Yang-Mills theory in the presence of strings cloud and quintessence can be found in \cite{Sadeghi:2022kgi}. The black holes with cloud of strings and quintessence in Lovelock gravity were studied in\cite{deMToledo:2018tjq}.\\
In this paper, we want to investigate the effects of  a collection of strings and quintessence on phase transition in Rastall gravity. Then we use the results of this research to explain the dual description of this event on CFT side.
\section{Black hole solutions in Rastall AdS  gravity  with the cloud of strings and quintessence}
\label{sec2}
Rastall's field equation in the presence of quintessence and cloud of strings in $d$ dimensions is as follows,
\begin{equation}\label{EoM}
	R_{\mu\nu}-\frac{1}{2}(R+\frac{d_1 d_2}{l^2})g_{\mu\nu}+{k} \lambda g_{\mu\nu}R=kT_{\mu \nu}.
\end{equation}
Here, quantities $R_{\mu\nu}$, $R$, $k$ , $l$ and $\lambda$ denote the Ricci tensor, Ricci scalar, Rastall gravitational coupling constant, AdS raduis and  Rastall parameter respectively. Where, $d_1$ and $d_2$  are related by $d_n=d-n$ and $T^{\mu \nu}$ is decomposed to the quintessence and cloud of strings terms as follows,
\begin{equation}
	T_{\mu \nu}=T^{(quin)}_{\mu \nu}+T^{(cs)}_{\mu \nu}.
\end{equation}
On the other hand, the Einstein and Rastall tensors are,
\begin{equation}\label{Ein}
	G_{\mu \nu }=R_{\mu \nu}-\frac{1}{2}Rg_{\mu \nu},
\end{equation}
and
\begin{equation}\label{Rast}
	H_{\mu \nu }=G_{\mu \nu }+{k} \lambda g_{\mu\nu} R.
\end{equation}
By substituting equations (\ref{Ein}) and (\ref{Rast}) into equation (\ref{EoM}), we get,
\begin{equation}\label{Rast2}
	H_{\mu \nu }-\frac{d_1 d_2}{2l^2} g_{\mu\nu}=kT_{\mu \nu}.
\end{equation}
The components of the energy-momentum tensor of the quintessence are given by\cite{Sadeghi:2020lfe},
\begin{equation}
	{T_t}{{\mkern 1mu} ^t} = {T_r}{{\mkern 1mu} ^r} = {\rho _q} =  \frac{{\omega \alpha {d_1}{d_2}}}{{2{r^{{d_1}(\omega  + 1)}}}},
\end{equation}
and
\begin{equation}
	{T_{{x_i}}}{{\mkern 1mu} ^{{x_i}}} =  - \frac{{{\rho _q}}}{{{d_2}}}({d_1}\omega  + 1),\qquad (i=2,...,d_2),
\end{equation}
The factor, $\alpha$ is the quintessence parameter related to the density of quintessence field ${\rho _q}$. The parameter $\omega $ is the barotropic index. The equation of state for quintessence is given by $p_q = \omega \rho _q$. The value of barotropic index is placed in the range  $ - 1 < \omega < {{ - 1} \over 3}$. Since, $\omega=\frac{\alpha}{d_1}$, we have $ - {d_1}<\alpha  < {{ - {d_1}} \over 3}$.
The components of energy-momentum tensor for the cloud of strings are as follows\cite{Letelier:1979ej},
\begin{equation}
	{T_t}{{\mkern 1mu} ^t} = {T_r}{{\mkern 1mu} ^r} =\frac{a}{r^{d_2}},
\end{equation}
and
\begin{equation}
	{T_{{x_i}}}{{\mkern 1mu} ^{{x_i}}} =  0,
\end{equation}
where, $a$ denotes the parameter of cloud of strings.\\
To obtain the black hole solutions, we use the ansatz
\begin{equation}\label{Metric}
	ds^{2} =-f(r)dt^{2} +\frac{dr^{2}}{f(r)} +r^2 d\Omega^2_{d_2}.
\end{equation}
The form of $f(r)$ in $d$-dimensions is as follows,
\begin{equation}\label{d4}
	f(r)=\kappa-\frac{2m}{r^{d_3}} -\frac{2 \Lambda }{d_1 d_2}r^2-\frac{\alpha}{r^{d_1\omega+d_3}}-\frac{2a}{d_2 r^{d_4}}.
\end{equation}
The event horizons are obtained through the solving of equation $f(r_h)=0$. This gives the integration constant $m$ in the form,
\begin{equation}\label{m}
	m=\frac{\kappa r_h^{d_3}}{2}-\frac{\Lambda }{d_1 d_2}r_h^{d_1}-\frac{\alpha}{2r_h^{d_1\omega}}-\frac{a}{d_2}r_h.
\end{equation}
The integration constant $m$ is related to the ADM mass of the black hole. For the metric in relation (\ref{d4}), the ADM mass is obtained by relation
\begin{equation}\label{adm}
	M=\left(\frac{d_2\Omega_{d_2}}{8\pi}\right)m,
\end{equation}
where, $d_n$ has been defined as $d_n=d-n$. In the study of thermodynamic properties of the black holes, the ADM mass of the black hole is considered as the enthalpy of the system. 
The Hawking temperature of the black hole is obtained through the relation
\begin{equation}\label{th}
	T=\frac{1}{4 \pi}\left(\frac{\partial f(r)}{\partial r}\right)_{r=r_h}.
\end{equation}
We shall use this relation in obtaining the entropy of the system through the first law of thermodynamics.\par
The entropy  and surface area of a black hole solution are given by relations\cite{Hawking1983},
\begin{equation}
	A=\int d^{d_2} x \sqrt{-g} |_{r=r_h,t=cte}= r_{h}^{d_2} \Omega_{d_2},
\end{equation}
and
\begin{equation}\label{ent}
	S=\frac{A}{4G} =\frac{r_{h}^{d_2} \Omega_{d_2} }{4 G}.
\end{equation}
Where, $\Omega_{d_2}$ is the surface area of the unit $d-2$ dimensional sphere which is given by,
\begin{equation}\label{area}
	\Omega_{d_2}=\frac{ 2 \pi^{\frac{d_1}{2}}}{\Gamma(\frac{d_1}{2})}
\end{equation}
The Entropy of the system for $d=4$($G=1$) is as follows,
\begin{equation}\label{s4}
	S=\frac{r_{h}^{2} \Omega_{2} }{4 G}=\pi r_h^2,
\end{equation}
while for $d=5$ takes the form,
\begin{equation}\label{s5}
	S=\frac{r_{h}^{3}\Omega_{3}}{4G}=\frac{\pi^2 r_h^3}{2}
\end{equation}
In the following, we study the thermodynamics of the system in four and five dimensions.
\section{The thermodynamics of 4-dimensional solution }
\label{sec4}
The ADM mass or enthalpy of the system in four dimensions, can be determined through the relations (\ref{m}) and (\ref{adm}). It is given by
\begin{equation}\label{mass1}
	M=m=\frac{r_h}{2}-\frac{\Lambda }{6}r_h^{3}-\frac{\alpha}{2r_h^{3\omega}}-\frac{a}{2}r_h.
\end{equation}
In addition to the relation (\ref{ent}), the entropy of the system can also be obtained through the first law of thermodynamics. In other words, we have
\begin{equation}\label{termo1}
	S=\int_{0}^{r_h} \frac{1}{T} \frac{\partial M}{\partial r_h} dr_h=\pi r_h^2,
\end{equation}
where, in which $T$ is determined by relation (\ref{th}). Now, relation (\ref{mass1}) can be written in terms of entropy and the pressure $P=-\frac{\Lambda}{8 \pi G}$ in four dimensions. The result is as follows,
\begin{equation}\label{mass2}
	M = \displaystyle -\frac{\pi^{\frac{3 \omega}{2}} (l^{2} S)^{-\frac{3 \omega}{2}} \alpha}{2}+\frac{4 \sqrt{l^{2} S}\, P \,l^{2} S}{3 \sqrt{\pi}}-\frac{\sqrt{l^{2} S}\, a}{2 \sqrt{\pi}}+\frac{\sqrt{l^{2} S}}{2 \sqrt{\pi}},
\end{equation}
where, we have assumed that $G=1$.
The first law of black hole ($\kappa=1$) thermodynamics in the extended phase space becomes\cite{Zou:2020jeu}-\cite{Grumiller:2022qhx}:
\begin{equation}
	dM=TdS+VdP +\mathcal{A} da+\mathcal{U}d \alpha,
\end{equation}
with
\begin{equation}\label{temp1}
	\begin{split}
		T&=\left(\frac{\partial M}{\partial S}\right)_{P,\alpha,a}\\
		&=\frac{3 \pi^{\frac{3 \omega}{2}} S^{-\frac{3 \omega}{2}-1} \omega  \alpha}{4 l^{3\omega}}+\frac{2 l^{3} \sqrt{S}\, P}{\sqrt{\pi}}-\frac{l a}{4 \sqrt{S}\, \sqrt{\pi}}+\frac{l}{4 \sqrt{S}\, \sqrt{\pi}},
	\end{split}
\end{equation}
\begin{equation}\label{vol}
	V=\left(\frac{\partial M}{\partial P}\right)_{S,\alpha,a}= \frac{4}{3\sqrt{\pi}} \left(l^2 S \right)^{\frac{3}{2}}=\frac{4\pi}{3}r_{h}^3,
\end{equation}
\begin{equation}
	\mathcal{A}=\left(\frac{\partial M}{\partial a}\right)_{S,P,\alpha}=-\frac{(l^2S)^{\frac{1}{2}}}{2\sqrt{\pi}}=-\frac{r_{h}}{2},
\end{equation}
and
\begin{equation}
	\mathcal{U}=\left(\frac{\partial M}{\partial \alpha}\right)_{S,P,a}=-\frac{1}{2}\pi^{\frac{3\omega}{2}}(l^2S)^{-\frac{3\omega}{2}}=-\frac{r_{h}^{-3\omega}}{2}.
\end{equation}
The temperature (\ref{temp1}) in terms of cosmological constant and black hole horizon changes to,
\begin{equation}\label{temp2}
	T=\frac{l^2}{4 \pi}\left(\frac{3  \alpha  \omega}{r_{h}^{3\omega+2} }- \Lambda r_{h}-\frac{ a}{  r_{h}}+\frac{1}{  r_{h}}\right).
\end{equation}
It is obvious that for the large values of horizon $r_h$, the temperature in relation (\ref{temp2}) tends to the temperature of the Ads-Schwarzschild black hole.
By using relations (\ref{temp1}) or (\ref{temp2}), the pressure of the black hole is obtained as follows,
\begin{equation}\label{p1}
	P=-\frac{3 r_{h}^{-3-3 \omega} \omega  \alpha}{8 \pi}+\frac{T}{2 r_{h} l^{2}}+\frac{a}{8 \pi  r_{h}^{2}}-\frac{1}{8 \pi  r_{h}^{2}}.
\end{equation}
We limit ourselves to the two important limits $\omega=-1$ and $\omega=-\frac{1}{3}$ of the barotropic index to investigate the Van der Waals-like behavior of the black hole. 
Explicitly, there is no Van der Waals-like behavior in our system in four dimensions. Because, the term $\frac{1}{V}(\propto \frac{1}{r_{h}^3})$ or its higher orders are not seen in the pressure relation (\ref{p1}). 
\par
\begin{figure}[h!]
	\centerline{\includegraphics[width=5cm]{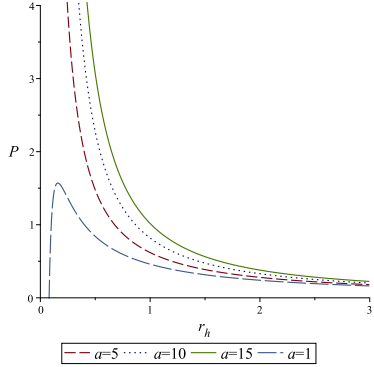}}
	\caption{$P-r_{h}$ diagram for $\omega=-\frac{1}{3}(\alpha=-1)$, and $T=1$ for different values of parameter $a$. No critical point is observed. \label{fig:1}}
\end{figure}

As usual, a critical point occurs when $P$ satisfies the conditions,
\begin{equation}\label{cr1}
	\frac{\partial P}{\partial r_h}|_{T=T_c,r_h=r_c}=0 \quad  and \quad \frac{\partial^2 P}{\partial^2 r_h}|_{T=T_c,r_h=r_c}=0,
\end{equation}
simultaneously \cite{Zou:2020jeu}. Our analytic and numerical investigations show that there is no critical point in $P-r_{h}$ plane in four dimensions. See the behavior of the black hole pressure in  
Fig (\ref{fig:1}) for the case $\omega=-\frac{1}{3}$.


\begin{figure}[h!]
	\centerline{\includegraphics[width=5cm]{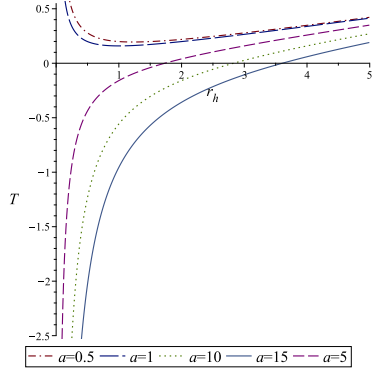}}
	\caption{$T-r_{h}$ diagram for $\omega=-\frac{1}{3} (\alpha=-1)$, $\Lambda=-1$ and different values of parameter $a$. \label{fig:2}}
\end{figure}

The behavior of the temperature of the system can be seen in Fig (\ref{fig:2}) for the case $\omega=-\frac{1}{3}$. 
\par
\begin{figure}[h!]
	\centerline{\includegraphics[width=5cm]{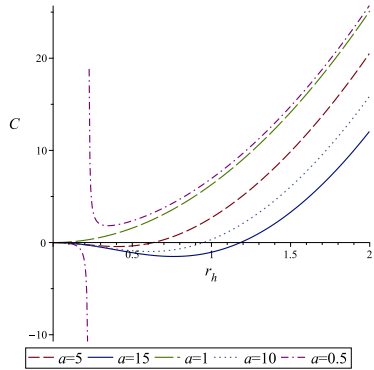}}
	\caption{$C-r_{h}$ diagram for $\omega=-1 (\alpha=-3)$, $\Lambda=-1$, $l=1$ and different values of $a$.  \label{fig:3}}
\end{figure}
\begin{figure}[h!]
	\centerline{\includegraphics[width=5cm]{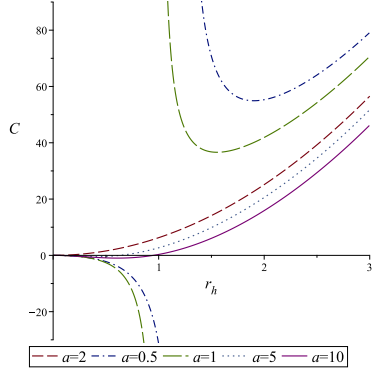}}
	\caption{$C-r_{h}$ diagram for $\omega=-\frac{1}{3} (\alpha=-1)$, $\Lambda=-1$, $l=1$ and different values of $a$. \label{fig:4}}
\end{figure}
\begin{figure}[h!]
	\centerline{\includegraphics[width=5cm]{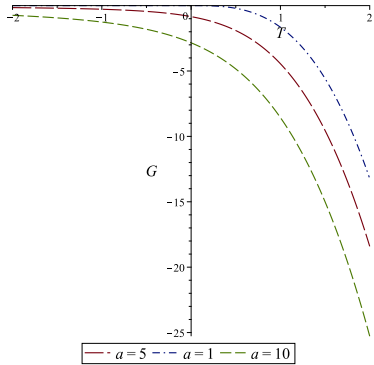}}
	\caption{$G-T$ diagrams for $\omega=-1 (\alpha=-3)$, $P=\frac{1}{8\pi}$ and several values of the parameter $a$. For $\omega=-\frac{1}{3} (\alpha=-1)$, a similar diagram is obtained. This shows that the Gibbs function is continuous against the temperature $T$ and there is no first-order transition in four dimensions. \label{fig:5}}
\end{figure}

The heat capacity of the black hole can be obtained through the relation $C=\frac{T}{\frac{\partial T}{\partial S}}$. It is given by,
\begin{equation}\label{C1}
	C=\frac{2 \pi  r_{h}^{2} \left(r_{h}^{1+3 \omega} a -r_{h}^{1+3 \omega} +r_{h}^{3 \omega +3} \Lambda -3 \omega  \alpha \right)}{l^{2} \left(9 \omega^{2} \alpha +r_{h}^{3 \omega +3} \Lambda -r_{h}^{1+3 \omega} a +r_{h}^{1+3 \omega} +6 \omega  \alpha \right)},
\end{equation}
in terms of cosmological constant and black hole horizon. We investigate the cases $\omega=-1$ and $\omega=-\frac{1}{3}$ which are of particular importance. By assuming that $\Lambda=-1$ and $l=1$ we have:
\begin{equation}\label{ch}
	C \vert_{\omega=-1}=-\frac{2\pi  r_{h}^{2} \left(-10 r_{h}^{2}+a -1 \right)}{10 r_{h}^{2}+a -1},
\end{equation}
for the maximum effect of quintessence and
\begin{equation}\label{cl}
	C \vert_{\omega=-\frac{1}{3}}=-\frac{2 \pi r_{h}^{2} \left(-r_{h}^{2}+a -2\right)}{r_{h}^{2}+a -2},
\end{equation}
for its minimum effect. The discontinuities of functions (\ref{ch}) and (\ref{cl}) are obtained by setting those denominators equal to zero. The results are $\left(\frac{\sqrt{-10 a+10}}{10},-\frac{\sqrt{-10 a+10}}{10}\right)$ for $C \vert_{\omega=-1}$ and $\left(\sqrt{-a+2},-\sqrt{-a+2}\right)$ for function $C \vert_{\omega=-\frac{1}{3}}$. The behavior of heat capacity functions (\ref{ch}) and (\ref{cl}) for different values of $a$ is seen in Figs (\ref{fig:3}) and (\ref{fig:4}). For the function $C \vert_{\omega=-1}$, the discontinuities occur for $a<1$. While, for  $C \vert_{\omega=-\frac{1}{3}}$, occur for $a<2$.
We found that there is no critical point for the pressure of the system in four dimensions and we shall see there is no first order transition for this case (the first derivative of the Gibbs free energy is continues everywhere.). On the other hand, the heat capacity which is related to the second derivative of the Gibbs function ($C=-T\frac{\partial^2 G}{\partial T^2}$), has discontinuities. In other words, the volume and the entropy change continuously and we have only second-order phase transition in four dimensions.
The Gibbs free energy is obtained through the relation,
\begin{equation}\label{G1}
	G=M-TS=\left(-\frac{3}{4} \alpha  \omega -\frac{1}{2} \alpha \right) r_{h}^{-3 \omega}+\frac{\Lambda  r_{h}^{3}}{12}+\left(\frac{1}{4}-\frac{a}{4}\right) r_{h}.
\end{equation}
These equations have complicated forms in terms of temperature, so we avoid writing them here. But, the investigation of Gibbs function (\ref{G1}) for two limits $\omega=-1$ and $\omega=-\frac{1}{3}$ shows that the first order phase transition does not occur in $G-T$ plane. 
The behavior of Gibbs function for the cases $\omega=-1$ and $\omega=-\frac{1}{3}$ are the same. The continues behavior of the Gibbs function for $\omega=-1(\alpha=-3)$, and different values of the parameter $a$ is seen in Fig (\ref{fig:5}).
\section{The thermodynamics of 5-dimensional solution}
\label{sec5}
In $d=5$ dimensions, the metric function $f(r)$ takes the form,
\begin{equation}\label{f5}
	f(r) = \kappa-\frac{2m}{r^{2}} -\frac{ \Lambda }{6}r^2-\frac{\alpha}{r^{4 \omega+2}}-\frac{2a}{3 r}.
\end{equation}
The enthalpy of the system is as follows,
\begin{equation}\label{m5}
	M =\frac{3 \pi}{4}m=\frac{3 \pi}{4}\left( \frac{\kappa r_h^{2}}{2}-\frac{\Lambda }{12}r_h^{4}-\frac{\alpha}{2r_h^{4\omega}}-\frac{a}{3}r_h\right).
\end{equation}
Now, the entropy of the system can  be obtained from the first law of thermodynamics. In other words,
\begin{equation}\label{termo2}
	S=\int_{0}^{r_h} \frac{1}{T} \frac{\partial M}{\partial r_h} dr_h=\frac{3 \pi}{4}\int_{0}^{r_h} \frac{1}{T} \frac{\partial m}{\partial r_h} dr_h=\frac{\pi^2 r_h^3}{2},
\end{equation}
where, $T$ is the Hawking temperature determined through the relation (\ref{th}).
This shows that the entropy from the first law of thermodynamics is the same as the entropy obtained from the area law of the black hole.  
The Smarr formula for $5$-dimensional case becomes:
\begin{equation}
	2 M=3 TS -2VP+\mathcal{A}a+\mathcal{U}\alpha.
\end{equation}
The associated thermodynamic quantities are as follows,
\begin{equation}\label{temp5}
	T =\frac{\,\pi^{\frac{8 \omega}{3}+1} 2^{-\frac{4 \omega}{3}} S^{-\frac{4 \omega}{3}} \omega  \alpha}{2 S}+\frac{\kappa  2^{\frac{2}{3}}}{4 S^{\frac{1}{3}} \pi^{\frac{1}{3}}}+\frac{4 S^{\frac{1}{3}} P 2^{\frac{1}{3}}}{3 \pi^{\frac{2}{3}}}-\frac{\pi^{\frac{1}{3}} a 2^{\frac{1}{3}}}{12 S^{\frac{2}{3}}}
\end{equation}
\begin{equation}\label{vol5}
	V=\left(\frac{\partial M}{\partial P}\right)_{S,\alpha,a}= \frac{2 \left(l^{3} S \right)^{\frac{1}{3}} 3^{\frac{1}{3}} l^{3} S}{\pi^{\frac{1}{3}}}=\frac{\pi^2  r_{h}^{4}}{2},
\end{equation}
\begin{equation}
	\mathcal{A}=\left(\frac{\partial M}{\partial a}\right)_{S,P,\alpha}=-\frac{3^{\frac{1}{3}} \left(l^{3} S \right)^{\frac{1}{3}}}{3 \pi^{\frac{1}{3}}}=-\frac{\pi r_{h}}{4},
\end{equation}
\begin{equation}
	\mathcal{U}=\left(\frac{\partial M}{\partial \alpha}\right)_{S,P,a}=-\frac{3^{1-\frac{4 \omega}{3}} \pi^{\frac{2}{3}+\frac{4 \omega}{3}} \left(l^{3} S \right)^{-\frac{4 \omega}{3}}}{6 \pi^{\frac{2}{3}}}=-\frac{3 \pi}{8 r_{h}^{4 \omega}}.
\end{equation}
The temperature in relation (\ref{temp5}), in terms of $\Lambda$ and black hole horizon is,
\begin{equation}\label{temp52}
	T=\frac{\alpha  \omega}{\pi  r_{h}^{3} \left(r_{h}^{\omega}\right)^{4}}-\frac{r_{h} \Lambda}{6 \pi}+\frac{\kappa}{2 \pi  r_{h}}-\frac{a}{6 \pi  r_{h}^{2}}
\end{equation}
where, gives the The black hole pressure as follows,
\begin{equation}\label{t52}
	P=\frac{3 T}{4 r_{h}}-\frac{3 \kappa}{8 \pi  r_{h}^{2}}+\frac{a}{8 \pi  r_{h}^{3}}-\frac{3 \alpha  \omega}{4 \pi  r_{h}^{4\omega+4}}.
\end{equation}
The pressure of the black hole($\kappa=1$) for the limits $\omega=-1 (\alpha=-4)$ and $\omega=-\frac{1}{3} (\alpha=-\frac{4}{3})$ and ($l=1$) are given by relations,
\begin{equation}\label{p5l}
	P\vert_{\omega=-\frac{1}{3}}=-\frac{3}{8 \pi  r_{h}^{2}}+\frac{3 T}{4 r_{h}}+\frac{a}{8 \pi  r_{h}^{3}}-\frac{1}{3 \pi  r_{h}^{\frac{8}{3}}},
\end{equation}
and
\begin{equation}\label{p5h}
	P\vert_{\omega=-1}=-\frac{3}{8 \pi  r_{h}^{2}}+\frac{3 T}{4r_{h}}+\frac{a}{8 \pi  r_{h}^{3}}-\frac{3}{\pi}.
\end{equation}
The heat capacity of the system is
\begin{equation}\label{c52}
	C=\frac{3 \pi^{2} r_{h}^{3} \left(\Lambda  r_{h}^{4 \omega +4}-3 r_{h}^{4 \omega +2} \kappa +r_{h}^{1+4 \omega} a -6 \omega  \alpha \right)}{2 \Lambda  r_{h}^{4 \omega +4}+6 r_{h}^{4 \omega +2} \kappa -4 r_{h}^{1+4 \omega} a +48 \alpha  \,\omega^{2}+36 \omega  \alpha},
\end{equation}
where, for two limits $\omega=-1 (\alpha=-4)$ and $\omega=-\frac{1}{3} (\alpha=-\frac{4}{3})$, takes the forms,
\begin{equation}\label{c5h}
	C \vert_{\omega=-1}=-\frac{3 \pi^{2} \left(-25 r_{h}^{3}+a -3 r_{h}\right) r_{h}^{3}}{50 r_{h}^{3}+4 a -6 r_{h}},
\end{equation}
and
\begin{equation}\label{c5l}
	C \vert_{\omega=-\frac{1}{3}}=-\frac{9 \pi^{2} r_{h}^{3} \left(3 r_{h}^{3}+9 r_{h}+8 r_{h}^{\frac{1}{3}}-3 a \right)}{-18 r_{h}^{3}+80 r_{h}^{\frac{1}{3}}-36 a +54 r_{h}}.
\end{equation}
The Gibbs free energy function in terms of  black hole pressure and its horizon is as follows,
\begin{equation}\label{gg5}
	G=-\frac{\pi  \alpha  \omega}{2 r_{h}^{4\omega}}-\frac{3 \pi  \alpha}{8r_{h}^{4\omega}}-\frac{\pi^{2} P r_{h}^{4}}{6}+\frac{\pi  \kappa  r_{h}^{2}}{8}-\frac{\pi  a r_{h}}{6}.
\end{equation}
Again, for the limits $\omega=-1 (\alpha=-4)$ and $\omega=-\frac{1}{3} (\alpha=-\frac{4}{3})$, we have,
\begin{equation}\label{g5h}
	G\vert_{\omega=-1}=-\frac{1}{2} \pi  r_{h}^{4}-\frac{1}{6} \pi^{2} P r_{h}^{4}+\frac{1}{8} \pi  r_{h}^{2}-\frac{1}{6} \pi  a r_{h},
\end{equation}
and
\begin{equation}\label{g5l}
	G\vert_{\omega=-\frac{1}{3}}=\frac{5 \pi  r_{h}^{\frac{4}{3}}}{18}-\frac{\pi^{2} P r_{h}^{4}}{6}+\frac{\pi  r_{h}^{2}}{8}-\frac{\pi  a r_{h}}{6}.
\end{equation}
\par

\begin{figure}[h!]
	\centerline{\includegraphics[width=5cm]{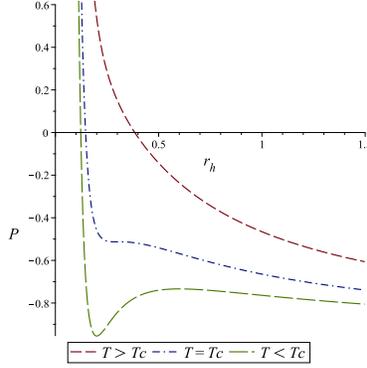}}
	\caption{$P-r_h$ diagram for $\omega=-1 (\alpha=-4)$, $\kappa=1$, $a=0.3$ and different values of $T$. \label{fig:6}}
\end{figure}

\begin{figure}[h!]
	\centerline{\includegraphics[width=5cm]{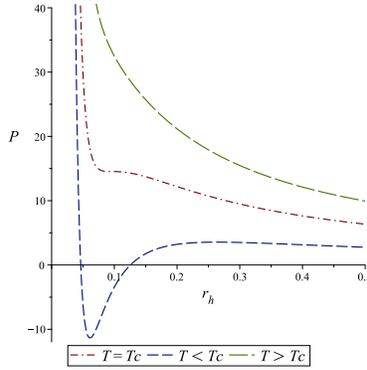}}
	\caption{$P-r_h$ diagram for $\omega=-\frac{1}{3} (\alpha=-\frac{4}{3})$, $\kappa=1$, $a=1$ and different values of $T$. \label{fig:7}}
\end{figure}

Now, to investigate the critical behavior of the system, we use the condition (\ref{cr1}). For $d=5$, we find critical point for the pressure of system. For $\omega=-1$, the critical quantities are as follows,
\begin{equation}\label{rc}
	r_c\vert_{\omega=-1}=\frac{a}{\kappa},
\end{equation}
\begin{equation}\label{tc}
	T_c\vert_{\omega=-1}=\frac{\kappa^2}{2\pi a},
\end{equation}
and
\begin{equation}\label{pc}
	P_c\vert_{\omega=-1}=\frac{\kappa^3}{8\pi a^2}-\frac{3}{\pi}.
\end{equation}
We see that the critical horizon is proportional to the parameter $a$. In other words, the critical volume increases by increasing the value of the parameter $a$. The critical temperature is proportional to the inverse of cloud of string parameter. While, the critical pressure is proportional to the inverse of square of the parameter $a$.
The critical relations for the limit $\omega=-\frac{1}{3}$ are too complicated. So, for this case, we only depict the associated diagrams. The existence of critical points is obvious from Figs (\ref{fig:6}) and (\ref{fig:7}).
\par
\begin{figure}[h!]
	\centerline{\includegraphics[width=5cm]{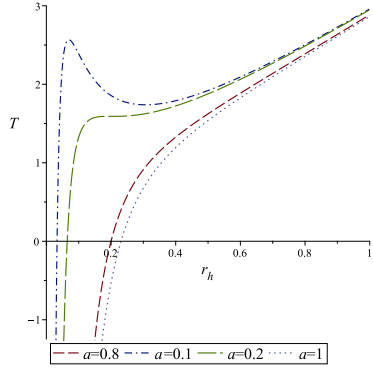}}
	\caption{$T-r_h$ diagram for $\omega=-1 (\alpha=-4)$,  $\kappa=1$, $\Lambda=-1$ and different values of $a$. \label{fig:8}}
\end{figure}
\begin{figure}[h!]
	\centerline{\includegraphics[width=5cm]{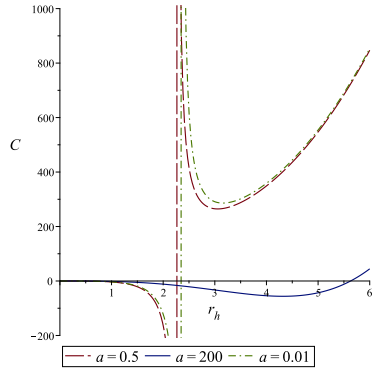}}
	\caption{$C-r_h$ diagrams for $\omega=-\frac{1}{3} (\alpha=-\frac{4}{3})$,  $\kappa=1$, $\Lambda=-1$ and different values of $a$.  \label{fig:9}}
\end{figure}
\begin{figure}[h!]
	\centerline{\includegraphics[width=5cm]{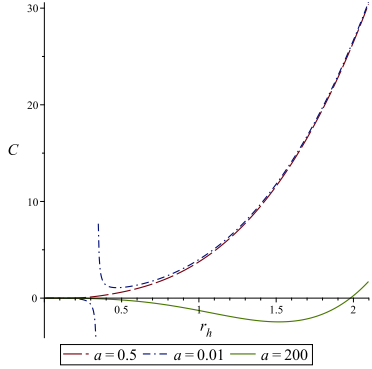}}
	\caption{$C-r_h$ diagrams for $\omega=-1 (\alpha=-4)$, $\kappa=1$, $\Lambda=1$ and different values of $a$. \label{fig:10}}
\end{figure}
\begin{figure}[h!]
	\centerline{\includegraphics[width=6cm]{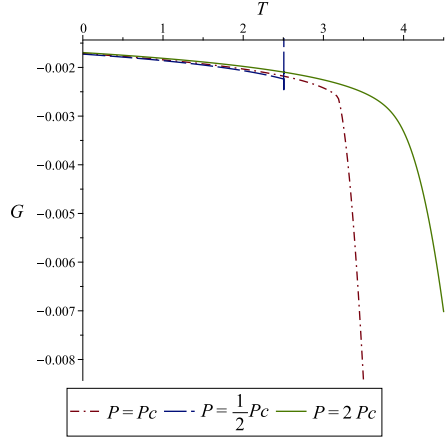}}
	\caption{$G-T$ diagrams for $\omega=-1 (\alpha=-4)$, $\kappa=1$, $a=0.1$ and different values of $P$. The discontinuity of the first derivative of the Gibbs function for $P=P_c$ is obvious in this figure. \label{fig:11}}
\end{figure}

The temperature of the system for the two important limits take the forms:
\begin{equation}\label{t5h}
	T\vert_{\omega=-1}=\frac{4 r_{h}}{\pi}-\frac{r_{h} \Lambda}{6 \pi}+\frac{1}{2 \pi  r_{h}}-\frac{a}{6 \pi  r_{h}^{2}},
\end{equation}
and
\begin{equation}\label{t5l}
	T\vert_{\omega=-\frac{1}{3}}=\frac{4}{9 \pi  r_{h}^{\frac{5}{3}}}-\frac{r_{h} \Lambda}{6 \pi}+\frac{1}{2 \pi  r_{h}}-\frac{a}{6 \pi  r_{h}^{2}}.
\end{equation}
The behavior of $5$-dimensional black hole temperature for the case $\omega=-1$ is seen in Fig (\ref{fig:8}) for different values of parameter $a$. The critical behavior can be also seen in this figure.
Now, we investigate the behavior of heat capacity of the system in five dimensions for two important cases $\omega=-1(\alpha=-4)$ and $\omega=-\frac{1}{3}(\alpha=-\frac{4}{3})$ and the different values of cloud of strings parameter.
The discontinuities in Figs (\ref{fig:9}) and (\ref{fig:10}) show that there is second order phase transition in this model for some values of cloud strings parameter. Because, as we mentioned before, the heat capacity is proportional to the second derivative of the Gibbs free energy function. However, we shall see that there is also the first order phase transition in this model. Black hole changes the phase between stable and unstable states. By increasing the parameter of cloud of strings, the black hole stability domain increases.
The investigation of the Gibbs free energy for the case $\omega=-1 (\alpha=-4)$ shows that the first-order phase transition occurs for this model in five dimensions. In other words, there is a discontinuity in the first derivative of the Gibbs free energy which shows a discontinues change in entropy of the system. We depict three diagrams with $P=\frac{1}{2}P_c$, $P=P_c(T=T_c)$ and $P=2P_c$. See Fig (\ref{fig:11}). For pressures grater than the critical pressure(relation (\ref{pc})), the discontinuity in the first derivative of the Gibbs free energy disappears. While for the values less than the critical value of the pressure, in addition to discontinuity in first derivative of the Gibbs function, the continuity of the function itself is lost. The discontinuity of the first derivative of the Gibbs function for $P=P_c$ is obvious from the Fig (\ref{fig:11}).
These investigations show that the confinement-confinement phase transition occurs only in five dimensions in this model.\\  
The Hawking-Page phase transition or the first-order phase transition is between the thermal AdS solution and a Schwarzschild black hole. There is the stable black hole solution above the critical temperature $T_c$, but there is the thermal AdS solution below the critical temperature $T_c$. The Hawking-Page phase transition means a confinement-deconfinement phase transition in the AdS/CFT correspondence. Quarks are confined to be grouped together in pairs or triples at temperatures below some critical temperature $T_c$, but they are in a deconfined phase at higher temperatures occurring freely in a quark-gluon plasma. Gubser \cite{Gubser:2005ih}-\cite{Gubser:2008px} shows that the second order phase transition corresponds to a phase transition from a normal phase to a superconducting phase in the dual field theory.
\section{Conclusion}
We introduced the Rastall AdS black hole solution with the cloud of strings and quintessence. We studied the phase transition of this model in four and five dimensions. Our outcomes show that in four dimensions only the second-order phase transition occurs. In other words, we observed the discontinuities in the heat capacity of the system which is related to the second derivative of the Gibbs function. While in five dimensions, the first-order transition is also seen. This means volume and entropy change discontinuously at critical points in five dimensions. According to AdS/CFT correspondence, our results indicate that the confinement-deconfinement phase transition occurs in the field theory side of this model in five dimensions.\\
As we know, the cosmological constant and the Rastall coupling are related as,
$\Lambda=\frac{1}{3} \frac{\rho_0}{4 k \lambda-2}$.
Thus, the effect of Rastall coupling is included in the cosmological constant which plays the role of the pressure of the system. In fact, there is no essential difference between the phase transitions in the Rastall gravity with the other phase transitions of the AdS black holes.  Nevertheless, there are differences in the dual interpretation. The coupling of the dual field of Rastall gravity is stronger than the coupling of the dual field theory of AdS black holes.  Because, the ratio of the shear viscosity to entropy density of this model is $\frac{\eta}{s} \le \frac{1}{4 \pi}$, which it has been shown in \cite{Sadeghi:2018vrf}. We note that the ratio of the shear viscosity to entropy density is proportional to the inverse of the square of the coupling of the dual field theory, $\frac{\eta}{s} \sim \frac{1}{\lambda^2}$.

\vspace{1cm}
\noindent {\large {\bf Acknowledgment} }   We would like to thank Juan Maldacena and David S. Berman for exchanging an email. We also thank the referees of IJMPA for the valuable comments which helped us to improve the paper.

\vspace{1cm}
\noindent {\large {\bf Data Availability Statement} }  All data that support the findings of this study are included within the article.

\end{document}